\documentstyle[amssymb,aps,prc]{revtex}
\begin{document}
\title{$^{11}Li$ stability and its single particle structure}
\author{T. N. Leite$^{1}$, N. Teruya$^{1}$, M. Kyotoku$^{1},$ and C. L. Lima$%
^{2}$}
\address{(1) Departamento de F\'{\i}sica, Universidade Federal da\\
Para\'{\i}ba, 58051-970 Jo\~{a}o Pessoa, Pb, Brazil}
\address{(2) Nuclear Theory and Elementary Particle Phenomenology Group\\
Instituto de F\'{\i}sica, Universidade de S\~{a}o Paulo,\\
C.P. 66318, 05315-970 S\~{a}o Paulo, SP, Brazil}
\date{\today }
\maketitle

\begin{abstract}
The stability of the $^{11}Li$ exotic nucleus is discussed focusing on the
existence of single particle positive energy components in its ground state
structure. The existence of these continuum components is an outcome of many
models using the available experimental data of $^{10}Li$. We show that if
this were the case, the $^{11}Li$\ main decay mode could not be the beta
decay.{\rm \ }
\end{abstract}


The structure of light exotic nuclei has been intensively investigated in
the last few years, both from experimental and theoretical points of view.
These nuclei are experimentally produced in laboratory using radiative beams
and in nature in the interior of exploding stars. A plethora of approaches
have been developed \cite{nois,broglia,OutrosEstrutura,3corpos} to describe
the paradigm of these exotic nuclei, the $^{11}Li$, a system composed by a $%
^{9}Li$ core plus a halo of two neutrons. This nucleus has remarkable
properties: due to the dilute characteristic of the valence nucleons wave
function, its radius is comparable to the $^{208}Pb$ ones; it has a
Borromean structure: if a halo neutron is taken away, its valence partner is
immediately emitted since the $^{10}Li$ is not bound. As a matter of fact, $%
^{11}Li$ decays mainly via beta emission, having a mean life of 8.5 ms. This
work intends to discuss the compatibility of the main decay mode of this
nucleus with the present knowledge about the structure of its ground state
wave function.

One of the many concerns regarding these dilute systems is the ordering of
the single particle states. The usual shell model, developed for the stable
nuclei, fails when applied to the fringes of the stability region. It has
been shown the importance of an adequate treatment of the residual
interaction of the valence particles among themselves and with the $^{9}Li$
core. \cite{nois,broglia,vinhmau}. At variance with the stable region, the $%
2s_{1/2}$ seems to be lower than the $1d_{5/2}$ and, according to some
experimental indications, even lower than the $1p_{1/2}$. Woods-Saxon
potentials with increased diffuseness are able to take into account the
first effect \cite{hamamoto} but for the second one the particle-vibration
coupling seems to be essential \cite{broglia,vinhmau}.

Many of the models proposed to describe the $^{11}Li$ deal primarily with
its ground state structure. Here, the development is strongly guided by the
experimental results; they often have indicated a couple of positive energy
single particle states in the $^{10}Li$ spectra, identified as $1p_{1/2}$
and $2s_{1/2}$ \cite{young+}. Based on these results, most of the
theoretical ground state wave functions are a model depend combination of
these two states. There is, however, a strong debate in the recent
literature regarding the nature of the valence states occupied by the halo
neutrons. In a recent paper, Chartier et al. \cite{chartier} by analyzing
the fragment-$\gamma $-ray coincidences from the breakup of $^{11}Be$\ were
able to assign the configuration $\pi 1p_{3/2}\otimes \nu 2s_{1/2}$\ to the
ground state of $^{10}Li$. Caggiano et al. \cite{caggiano} have obtained
different results; they reported measurements of the low-lying resonance
structure in $^{10}Li$\ aiming at a clean, high resolution, and high
statistic spectroscopic determination of the $^{10}Li$\ spectra near
threshold. Their result is somehow surprising: they could not see any
indication of a $2s_{1/2}$\ state. Another experimental paper \cite{betadec}%
, however, has remeasured the $\gamma $ ray intensities following the $\beta 
$ decay of $^{11}Li$ and claims that a ground state wave function containing
about 50\% of $s$-wave neutron component is necessary in order to get
agreement between theory and experiment, for both the half life and
branching ratio of the $^{11}Li$ decay to the first excited state of $%
^{11}Be $.

The present experimental uncertainties regarding the real nature of the
single particle structure of the exotic nuclei in the $Li$ region poses a
rather difficult problem to any attempt to a theoretical understanding of
the ground state structure of $^{11}Li$. Aditionally, on the theoretical
arena, the situation is not anyhow better. Masui et al. \cite{masui}
analyzed the $^{9}Li+n$ elastic scattering $s$-wave cross sections and
concluded that there is no low-lying $s$-wave resonant state but rather a
``virtual'' one, whereas Caurier et al. \cite{caurier} performed shell model
studies on neutron rich nuclei and predicted a ground state of $^{11}Li$
dominated by a configuration with two neutrons in the $2s_{1/2}$ orbit.

The present note discuss another aspect of this problem, namely, the width
of the single particle resonances and its relation with the decay modes of $%
^{11}Li$. Since the $^{11}Li$ decays through beta emission,\ any width
associated to the particle channel should necessarily be smaller than $\sim
5\times 10^{-17}$ KeV in order to keep the beta emission as the main decay
mode. Many of the theoretical descriptions of the $^{11}Li$, however,\ have
as an outcome at least one continuum single-particle component in the
ground-state wave function; this is necessary in order to reproduce both the
two-neutron binding energy and the halo radius. At this point, we would like
to mention that the existence of a continuum component does not necessarily
imply in a direct neutron decay\ and the neutrons could be bound for a time
long enough to allow the nucleus still to be a beta emitter, depending on
the width associated with each decay process. The main question posed by the
present work is to discuss compatibility of the requirement regarding the
main decay mode with a ground state wave function with a component in the
continuum.

Our analysis of this problem will be performed in the framework proposed in
Ref.\cite{nois}; our conclusions, however, will not be limited by this
particular approach. In that work the BCS treatment of the residual pairing
interaction between the valence nucleons was used in such a way to include
explicitly the contribution of the residual interaction on the single
neutron energies. In Ref.\cite{nois} the mean value of a Bogoliubov-Valatin
transformed pairing Hamiltonian was written as: 
\begin{equation}
h=2\sum_{j}\left( \varepsilon _{j}-\lambda -\frac{G}{2}\right)
v_{j}^{2}-G\sum_{ij}\Omega _{j}\left( \Omega _{j}-\delta _{ij}\right)
u_{i}v_{i}u_{j}v_{j}\text{.}  \label{eq1}
\end{equation}%
In the above expression, $\varepsilon _{j}$\ is the single particle energy
of a state with angular momentum $j$\ ($2\Omega _{j}=2j+1$), $\lambda \ $is
the chemical\ potential, and $G$\ is the pairing strength; $v_{j}^{2}$\ is
the usual pair occupancy ($u_{j}^{2}=1-v_{j}^{2}$). The self-energy term, $%
\frac{G}{2}v_{j}^{2}$, in Eq. (\ref{eq1}) is a real renormalization arising
from the interacting valence particles, shifting the mean field at once. At
this point we would like to stress that each external neutron in $^{11}Li$\
has the redefined energy $\tilde{\varepsilon}_{j}=\varepsilon _{j}-\frac{G}{2%
}$, while in $^{10}Li$\ the odd neutron energy remains $\varepsilon _{j}$\
since the mean field does not change.

In Ref.\cite{nois} this model was applied to describe the $^{11}$Li ground
state.\ Just to show the feasibility of the approach, only a $1p_{1/2}$\
state was initially considered; in a second step, both $1p_{1/2}\ $($%
\varepsilon _{1p_{1/2}}=0.50$\ MeV) and $2s_{1/2}$\ ($\varepsilon
_{2s_{1/2}}=0.10$\ MeV) were taken into account. As discussed above, it was
essential the careful treatment of the self-energy contribution (this term
is usually neglected with the argument that its only effect is to
renormalize the single particle energies, which are in general extracted
from, or fitted to experimental data in neighbor nuclei). As a result, the
mean field felt by valence neutrons is more attractive in the $^{11}Li$\
than in the $^{10}Li$, due to the contribution of the residual pairing
interaction \cite{comment1}. In this way, $^{11}Li$\ may be bound contrary
to $^{10}Li$, which remains unbound. Even after this renormalization, the $%
1p_{1/2}$\ single particle state still lied in the continuum whereas the $%
2s_{1/2}$\ dived into the discrete region. The obtained ground state wave
function had quasi-particle occupancies, $v_{j}^{2},$\ equal to $0.19$\ and $%
0.81$\ for $1p_{1/2}$\ and $2s_{1/2}$, respectively. The corresponding
renormalized single particle energies were $\tilde{\varepsilon}%
_{2s_{1/2}}=-0.22$\ MeV and $\tilde{\varepsilon}_{1p_{1/2}}=0.18$\ MeV. At
this point, we emphasize that the two-neutron binding energy is mainly
granted by the $2s_{1/2}$\ state and the adopted value for $G$\ ($G=7/A=0.32$%
\ MeV) was not enough to allow the $1p_{1/2}$\ state to leave the continuum
region.\ We would like to call the reader's attention to the fact that if
only one single particle state were considered (irrespective of $s$ or $p$
state), this renormalization would have brought it to the discrete region
because a bigger value of $G\ $would\ be necessary to obtain the correct
binding energy. If this were the case, there would be no difficulties
regarding the compatibility between the wave function structure and the main
decay mode of $^{11}Li,$\ since the valence neutron would occupy a bound
energy level.

In a light system the low energy neutron resonance has a half-life as short
as $\sim 10^{-21}$ sec because the only barrier to hold the neutron inside
the well is the centrifugal one (with low orbital angular momentum).\ In $%
^{11}Li$, the $1p_{1/2}$\ state lying in continuum would be the main
responsible for a undesirable decay through particle emission. It is
therefore worth to analyze the width associated to this state.

According to the above discussion, the single particle energies in $^{11}Li$
have to be obtained through a potential with parameters different from those
which reproduce the single particle resonances in $^{10}Li$. Thereby, our
particular interest is to analyze the modifications on the widths of these
resonances, or more precisely if the redefinitions in energy and width would
stabilize the continuum component in the ground state wave function. We
begin on by studying the behavior of the $1n+^{9}Li$\ system. A spherical
Woods-Saxon potential was used to generate the scattering states available
to the neutron, calculated using the model of Ref. \cite{NT,Telio}, and the
results are showed in Table \ref{tab1} for three different sets of
Woods-Saxon parameters, denoted by PI, PII, and PIII. The parameters of the
potential PI were obtained in an attempt to reproduce the experimental
values of energy and width of the $\frac{1}{2}^{-}$\ resonance ($\varepsilon
_{1p_{1/2}}\thicksim 0.50$ MeV, $\Gamma _{1p_{1/2}}\thicksim 0.40$ MeV).
This resonance can be interpreted as a one-neutron state in an ordinary
Woods-Saxon potential with radius $R=1.16A^{1/3}$. In $^{11}Li$ the
situation is quite different because the external two halo neutrons have
special features: they are away from the core and they have a very low
binding energy. Therefore, as a next step, we try to reproduce the single
particle energy from a calculation able to obtain the ground state energy of 
$^{11}Li$. The parameters of the potential PII in Table \ref{tab1} were then
chosen in such a way to reproduce the redefined energy $\tilde{\varepsilon}%
_{1p_{1/2}}$ from Ref. \cite{nois}. We can see that the obtained potential
well is shallower than in the previous case. This behavior reproduces the
fact that in $^{11}Li$\ we have actually two different neutron categories:
the core neutrons and the halo ones. This $^{11}Li$ description is
compatible with the two oscillators model of Kuo et al. \cite{Kuo}. Although
the $1p_{1/2}$ state becomes more bound in $^{11}Li$ than in $^{10}Li$, it
still remains in the continuum. The adjusted potential gives rise to a very
broad resonance, which is unable to keep the two halo neutrons bound, even
for a short period of time. As a matter of fact, the continuum component of
the ground state destabilize the ground state. We are faced with a
contradiction since a ground state wave function, capable to reproduce
adequately the two neutrons binding energy, should not have a so broad
continuum component. This problem could be thought as a particular
difficulty of the model proposed in Ref. \cite{nois}. However, any
theoretical description of $^{11}Li$\ which includes a continuum component $%
(1p_{1/2}$\ or $2s_{1/2})$\ in the ground state wave function, suffers the
same disease.

The Woods-Saxon potential PIII in Table \ref{tab1} presents another aspect
of the above discussion. There, a $1p_{1/2}$\ resonant state with an
arbitrary and small positive redefined energy ($\tilde{\varepsilon}%
_{1p_{1/2}}=2$\ KeV) was constructed. This fictitious resonance would be
obtained, in the framework of Ref. \cite{nois}, by using a $G$\ pairing
strength adjusted in such a way to give an artificially big two-neutron
binding energy. In this case the resonance width is only $0.15$\ KeV broad,
still large enough to allow the ground-state to become neutron unstable. As
a remark, we would like to point out that some proton emitters have a decay
mechanism involving a competition between beta and proton channels. Even in
this case, the proton decay is the most likely \cite{batchelder}, indicating
that the continuum proton wave function destabilizes the nucleus before the
beta decay.

With regard to the $\frac{1}{2}^{+}$ resonance the situation is even more
critical because the $s$-wave does not have any centrifugal barrier to
overcome.\ Any positive energy would allow it quickly to escape, giving rise
to a very broad resonance making difficult its experimental determination %
\cite{Telio}.

In spite of the enormous uncertainties about the $^{10}Li$ structure, the
stability of the $^{11}Li$\ ground state against particle emission{\rm \ }%
can help us to estimate of the $\frac{1}{2}^{+}$\ resonance energy. By
keeping the $\frac{1}{2}^{-}$\ resonance at the accepted energy and adopting
a $G$\ value of the pairing strength in such a way to dive this resonant
state to the discrete energy region ($\varepsilon _{1p_{1/2}}=0.50$ MeV; $%
G=1.1$ MeV; $\tilde{\varepsilon}_{1p_{1/2}}=-0.05$ MeV), we can vary the
energy of the $\frac{1}{2}^{+}$ resonance with the constraint that the two
neutrons binding energy remains in a $400-1200$ KeV interval (we would like
to point out that: a) for the above value of $\varepsilon _{1p_{1/2}}$, any $%
G>1.0$\ MeV would produce a negative $\tilde{\varepsilon}_{1p_{1/2}}$;\ b) a
smaller value of $\varepsilon _{1p_{1/2}}$\ would be able to produce a two
neutrons separation energy interval closer to the experimentally accepted
value, namely, $294$ KeV. In any case, the conclusions below would remain
unchanged). This simple attempt to obtain a ground-state wave function whose
energy is around the expected value, with the proviso that it has to be
stable against particle emission - meaning a wave function without any
component in the continuum -, shows that an ``optimal'' value for the energy
of the $\frac{1}{2}^{+}$\ resonance is near the $\frac{1}{2}^{-}$\ one (see
Fig \ref{fig1}). This may perhaps explain the reason why it has been so
difficult to determine the real structure of the $^{11}Li$ ground state.

The main result of this work is to argue that no sensible description of
light halo nuclei can be achieved if their ground state wave function
contains any neutron component in the continuum. If this were the case, $%
^{11}Li$\ would decay through neutron emission instead of via beta decay,
since any positive energy component in the $^{11}Li$\ wave function prevents
this nucleus to be stable against neutron emission. This is a rather
stringent constraint on any model aiming at a description of these nuclei
and claims for an improvement of the present experimental knowledge of the
mean field near the drip line, in particular the status of the $\frac{1}{2}%
^{+}$\ state in $^{10}Li$.

{\bf Acknowledgement }NT and TNL are partially supported by the Conselho
Nacional de Desenvolvimento Cient\'{\i}fico e Tecnol\'{o}gico (CNPq).

%TCIMACRO{\TeXButton{B}{\begin{table}[tbp] \centering}}%
%BeginExpansion
\begin{table}[tbp] \centering%
%EndExpansion
\begin{tabular}{llllll}
$^{10}Li(1p_{1/2})$ & $V_{0}$ (MeV) & $R_{0}$ (fm) & $\varepsilon $ (KeV) & $%
\Gamma $ (KeV) & $T_{1/2}$ ($10^{-21}$s) \\ 
PI & \multicolumn{1}{c}{$36.2$} & \multicolumn{1}{c}{$2.5$} & 
\multicolumn{1}{c}{$508$} & \multicolumn{1}{c}{$339$} & \multicolumn{1}{c}{$%
\sim 1$} \\ 
PII & \multicolumn{1}{c}{$7.49$} & \multicolumn{1}{c}{$6.0$} & 
\multicolumn{1}{c}{$180$} & \multicolumn{1}{c}{$90$} & \multicolumn{1}{c}{$%
\sim 5$} \\ 
PIII & \multicolumn{1}{c}{$7.95$} & \multicolumn{1}{c}{$6.0$} & 
\multicolumn{1}{c}{$2$} & \multicolumn{1}{c}{$0.15$} & \multicolumn{1}{c}{$%
\sim 3\times 10^{3}$}%
\end{tabular}
\caption{The single-particle resonances for unbound $^{10}Li$ are presented.
 The Woods-Saxon potential parameters are adjusted in order to to reproduce:
 a) the experimental data (PI), b) the $\varepsilon _{1p_{1/2}}$ energy redefined by 
the pairing self-energy (PII), and c)  a potential producing an arbitrary energy state 
near zero (PIII).\label{tab1}}%
%TCIMACRO{\TeXButton{E}{\end{table}}}%
%BeginExpansion
\end{table}%
%EndExpansion

%TCIMACRO{\TeXButton{B}{\begin{figure}[tbp] \centering}}%
%BeginExpansion
\begin{figure}[tbp] \centering%
%EndExpansion
\caption{The energy of the $\frac{1}{2}^{+}$ resonance for unbound $^{10}Li$ is
estimated by taking into account stability against neutron emission in the $^{11}Li$ ground-state wave function. The pairing force strength, $G=1.1$
MeV, is chosen in such a way that the $\frac{1}{2}^{-}$ resonance ($\varepsilon _{1/2^{-}}=0.500$ MeV) dives into the discrete region ($\tilde{\varepsilon}_{1/2^{-}}=-0.050$ MeV) due to the effect of the pairing
self-energy. The renormalized energy of the $\frac{1}{2}^{+}$ resonance, $\tilde{\varepsilon}_{1/2^{+}}$, and the two neutrons binding energy, $\varepsilon _{2n}=\left( \varepsilon _{1/2^{+}}+\varepsilon
_{1/2^{-}}\right) -\frac{\left( \varepsilon _{1/2^{+}}-\varepsilon
_{1/2^{-}}\right) ^{2}}{2G}-\frac{3G}{2}$, are calculated by varying the
energy $\varepsilon _{1/2^{+}}$ and keeping the $\frac{1}{2}^{-}$ resonance
fixed at $\varepsilon _{1/2^{-}}=0.500$ MeV.
\label{fig1}}%
%TCIMACRO{\TeXButton{E}{\end{figure}}}%
%BeginExpansion
\end{figure}%
%EndExpansion


\begin{references}
\bibitem{comment1} Similar result was obtained in Ref. \cite{broglia} in
spite of a different treatment of the residual pairing interaction.
\end{references}
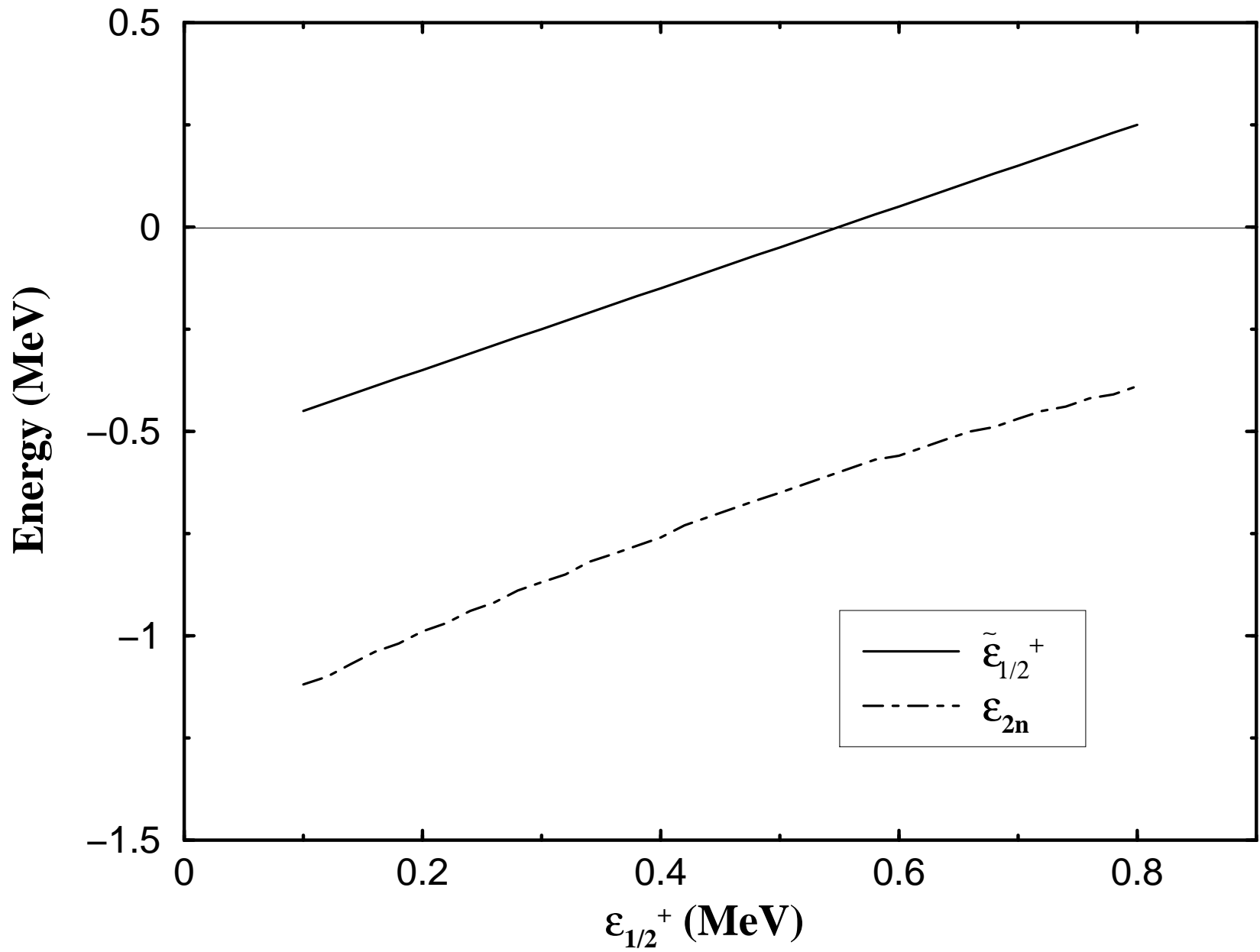
\end{document}